\documentclass[10pt,final,journal,letterpaper,twoside,twocolumn]{IEEEtran}
\usepackage{amsmath}
\usepackage{amssymb}
\usepackage{amsfonts}
\usepackage{amscd}
\usepackage{mathrsfs}
\usepackage{graphicx}
\usepackage{color}
\usepackage{url}
\usepackage{bm}
\usepackage{mdframed}

\DeclareMathOperator*{\argmin}{arg\,min}
\DeclareMathOperator*{\argmax}{arg\,max}

\newtheorem{theorem}{Theorem}

\begin{document}
\title{A Novel Power Allocation Scheme for Two-User GMAC with Finite Input Constellations}
\author{J. Harshan, \emph{Member, IEEE}, and B. Sundar Rajan, \emph{Senior Member, IEEE}
\thanks{Part of this work is in the proceedings of IEEE International Symposium on Personal, Indoor and Mobile Radio Communications (PIMRC 2011) held at Toronto, Canada, 11-14 September, 2011.}
\thanks{J. Harshan is with the Dept. of ECSE, Monash University, Australia. B. Sundar Rajan is with the Dept. of ECE, Indian Institute of Science, India. Email: harshan.jagadeesh@monash.edu, bsrajan@ece.iisc.ernet.in.}
}


\maketitle

\begin{abstract}
Constellation Constrained (CC) capacity regions of two-user Gaussian Multiple Access Channels (GMAC) have been recently reported, wherein an appropriate angle of rotation between the constellations of the two users is shown to enlarge the CC capacity region. We refer to such a scheme as the Constellation Rotation (CR) scheme. In this paper, we propose a novel scheme called the Constellation Power Allocation (CPA) scheme, wherein the instantaneous transmit power of the two users are varied by maintaining their average power constraints. We show that the CPA scheme offers CC sum capacities equal (at low SNR values) or close (at high SNR values) to those offered by the CR scheme with reduced decoding complexity for QAM constellations.
We study the robustness of the CPA scheme for random phase offsets in the channel and unequal average power constraints for the two users. With random phase offsets in the channel, we show that the CC sum capacity offered by the CPA scheme is more than the CR scheme at high SNR values. With unequal average power constraints, we show that the CPA scheme provides maximum gain when the power levels are close, and the advantage diminishes with the increase in the power difference.
\end{abstract}

\section{Introduction and Preliminaries}
\label{sec1}

In traditional networks, information exchange between the subscriber devices and the base station is realized by scheduling the transmissions in disjoint set of point-point channels, i.e., by \emph{separating} the subscribers through TDMA, FDMA, or CDMA. Over the last few decades, enormous research has taken place to look beyond \emph{channel separation}, and develop advanced physical layer techniques that can provide capacity gains at the cost of joint-processing of the signals from different subscribers \cite{Ga}-\cite{TsH2}. Along that direction, determining the capacity regions of multi-terminal networks \cite{CoT}, and designing appropriate low-complexity signalling schemes \cite{MudV} has received a lot of attention.

Till date, capacity regions are known only for a certain class of network configurations such as the multiple access channels (MAC) \cite{As, Li}, and Gaussian broadcast channels (BC) \cite{HYS}, to name a few. Though, contributions on such configurations started for channels with few users (such as two-user MAC/BC or three-user MAC/BC), the results have now been generalized to arbitrary number of users \cite{TsH1, TsH2, ASNS}. Further, for Gaussian MAC (GMAC), the capacity achieving input distribution is known to be Gaussian, which is a continuous distribution \cite{CoT}. However, in practice, the input constellations are finite in size and are uniformly distributed. As a result, the known results for continuous input does not shed light on the actual performance of finite constellations. This difference in the nature of the input alphabet has motivated researchers to revisit the GMAC (and similar channels), and study them from the view point of finite input constellations. As a first step, some results have been reported for channels with few users \cite{FTL}-\cite{RaR}. This paper is along that direction, and we deal with two-user GMAC with finite input constellations.

\indent Influenced by some preliminary works on MAC with finite inputs constellations \cite{FTL, FAR}, a detailed study on the CC capacity \cite{Eb} regions of two-user GMAC has been reported in \cite{HaR_TIT}. It is shown in \cite{HaR_TIT} that introducing an appropriate angle of rotation between the constellations provides enlargement in CC capacity region. Further, angles of rotation which maximizes the CC sum capacity have been provided for some known constellations. We refer to such a method of enlarging the CC capacity region as the Constellation Rotation (CR) scheme. Note that the CR scheme is a Non-Orthogonal Multiple Access (NO-MA) scheme, wherein the two users transmit during the same time and in the same bandwidth. It is also shown in \cite{HaR_TIT} that the CC capacity region of the CR scheme strictly encloses the CC capacity region of the FDMA and the TDMA. It is highlighted in \cite{HaR_TIT} that the above behaviour is not observed with Gaussian inputs, which in turn shows the importance of studying these channels with finite input constellations. Other than two-user GMAC, similar research on finite input constellations have also been reported for MAC with quantization and fading \cite{HaR_PIMRC}-\cite{MWC}, Gaussian broadcast channels \cite{NaS}, interference channels \cite{FrA, AbR}, relay channels \cite{WMCJ, WCMJ}, point-point MIMO channels \cite{WXDZS}, and secrecy channels \cite{RaR}. 

In the schemes proposed in \cite{HaR_TIT}-\cite{RaR}, either the channels are assumed to be fixed, or some form of channel state information is available at the transmitters. For instance, the CR scheme \cite{HaR_TIT} assumes fixed channels, and the technique is sensitive to the choice of the relative angle of rotation. If the channels introduce random phase offsets (say, due to clock synchronization problems), then the resultant relative angle need not provide maximum gains. In such a case, even if the random phase offsets are made available to the receiver, the CR scheme becomes ineffective as the transmitters do not have the knowledge of the phase values. As a result, there is a need for signal design which is robust to random phase offsets. In this paper, we propose a NO-MA scheme called the Constellation Power Allocation (CPA) scheme \cite{HaR_PIMRC}, wherein instead of introducing rotation between the constellations, we vary the transmit powers of the two users. Unlike the CR scheme, the proposed CPA scheme is robust to random phase offsets in the channel. The contributions of this paper are summarized as below:
\begin{itemize}
\item We propose a novel transmitter side technique called the CPA scheme to obtain enlargement in the CC capacity region of two-user GMAC. In the proposed scheme, the transmit power of the two users are varied while retaining the average power constraint for each user.
\item For the CPA scheme, the transmit power of the users are varied through a scale factor $\alpha \in [0, 1]$. For such a model, we propose the problem of finding an appropriate $\alpha$ that maximizes the CC sum capacity. For simpler computations, we propose a deterministic metric to compute the scale factor such that the CC sum capacity is maximized at high Signal-to-Noise Ratio (SNR) values. We compute the CC sum capacities for some known constellations, and show that the CC sum capacities offered by the CPA scheme are equal (at low SNR values) or close (at high SNR values) to those offered by the CR scheme in \cite{HaR_TIT}. (Section \ref{sec3})
\item For regular QAM constellations, we first identify that the CPA scheme provides a sum constellation $\mathcal{S}_{sum}$ whose in-phase and the quadrature components are separable, and subsequently show that the CPA scheme provides lower decoding complexity than the CR scheme. This advantage is shown to come with no significant reduction in the CC sum capacity. To exploit the reduced decoding complexity offered by the CPA scheme, we propose independent coding along the in-phase and the quadrature components for each user. We also propose TCM based code pairs to approach the CC sum capacity for 16-QAM constellation. We point out that the low decoding complexity advantages of the CPA are applicable only for GMAC with no random phase offsets. (Section \ref{sec4})
\item We study the robustness of the CPA scheme for two-user GMAC with random phase offsets in the channel. For such channels, it is clear that the CR scheme does not improve the CC sum capacity due to the random phase offsets unknown to the transmitters. We show that the CC sum capacity offered by the CPA scheme is more than the CR scheme at high SNR values. We also study the robustness of the CPA scheme for unequal average power constraints for the two users, where we show that the gains of the CPA scheme decreases as the power difference between the users increase. (Section \ref{sec5})
\end{itemize}


\textit{\textbf{Notations}:} Throughout the paper, boldface letters and capital boldface letters are used to represent vectors and matrices, respectively. For a random variable $X$ which takes value from the set $\mathcal{S}$, we assume some ordering of its elements and use $X(i)$ to represent the $i$-th element of $\mathcal{S}$, i.e., $X(i)$ represents a realization of the random variable $X$. We use the symbol $\imath$ to represent $\sqrt{-1}.$ Cardinality of a set $\mathcal{S}$ is denoted by $|\mathcal{S}|$. Absolute value of a complex number $x$ is denoted by $|x|$, and $E \left[x\right]$ denotes the expectation of the random variable $x$. A circularly symmetric complex Gaussian random vector, $\textbf{x}$ with mean $\bm{\mu}$ and covariance matrix $\mathbf{\Gamma}$ is denoted by $\textbf{x} \sim \mathcal{CSCG} \left(\bm{\mu}, \mathbf{\Gamma} \right)$.	

\section{Two-user GMAC: Signal Model and CC Sum Capacity}
\label{sec2}
The model of two-user GMAC consists of two users that need to convey independent information to a single destination. It is assumed that User-$1$ and User-$2$ communicate to the destination at the same time and in the same frequency band (the two users employ a NO-MA scheme). Symbol level synchronization is assumed at the destination. In this section, we assume no random phase offsets introduced by the channel.
The two users are equipped with \emph{finite complex} constellations $\mathcal{S}_{1}$ and $\mathcal{S}_{2}$ of size $N_{1}$ and $N_{2}$, respectively such that for $x_{i} \in \mathcal{S}_{i}$, we have  $E[|x_{i}|^{2}] = 1$. Let $P_{i}$ denote the average power constraint for User-$i$. When User-$1$ and User-$2$ transmit symbols $\sqrt{P_{1}}x_{1}$ and $\sqrt{P_{2}}x_{2}$ simultaneously, the destination receives the complex symbol $y$ given by
\begin{equation}
\label{gmac_modeleq}
y = \sqrt{P_{1}} x_{1} + \sqrt{P_{2}} x_{2} + z, ~~ \mbox{where }  z \sim \mathcal{CSCG} \left(0, \sigma^{2} \right),
\end{equation}
and $\frac{\sigma^{2}}{2}$ is the variance of the AWGN in each dimension.
The CC sum capacity of two-user GMAC \cite{HaR_TIT} is $I\left(\sqrt{P_{1}}x_{1} + \sqrt{P_{2}}x_{2} : y \right),$ which is given
\begin{figure*}
{\footnotesize
\begin{equation}
I(\sqrt{P_{1}}x_{1} + \sqrt{P_{2}}x_{2} : y) = \mbox{log}_{2}(N_{1}N_{2}) - \nonumber
\end{equation}
\begin{equation}
\label{mi}
\frac{1}{N_{1}N_{2}}\sum_{k_{1} = 0}^{N_{1} - 1}\sum_{k_{2} = 0}^{N_{2} - 1}E\left[\mbox{log}_{2}\left[ \frac{\sum_{i_{1} = 0}^{N_{1} - 1}\sum_{i_{2} = 0}^{N_{2} - 1} \mbox{exp}\left(- |\sqrt{P_{1}}x_{1}(k_{1}) + \sqrt{P_{2}}x_{2}(k_{2}) - \sqrt{P_{1}}x_{1}(i_{1}) - \sqrt{P_{2}}x_{2}(i_{2}) + z|^{2}/\sigma^{2}\right)}{\mbox{exp}\left(- |z|^{2}/\sigma^{2}\right)}\right] \right]
\end{equation}
}
\end{figure*}
in \eqref{mi} at the top of the next page.

\indent In \cite{HaR_TIT}, 
an appropriate angle of rotation between the constellations is shown to increase the CC sum capacity. In this paper, we introduce the CPA scheme to increase the CC sum capacity. Before introducing the CPA scheme, we recall the definition of the sum constellation and uniquely decodable (UD) constellation pairs \cite{HaR_TIT}. Given two constellations $\mathcal{S}_{1}$ and $\mathcal{S}_{2}$, the sum constellation of $\mathcal{S}_{1}$ and $\mathcal{S}_{2}$ is given by
\begin{equation*}
\mathcal{S}_{sum} \triangleq \left\lbrace \sqrt{P_{1}} x_{1} + \sqrt{P_{2}} x_{2} ~ | ~ \forall ~ x_{1} \in \mathcal{S}_{1}, x_{2} \in \mathcal{S}_{2}\right\rbrace.
\end{equation*}
The adder channel in the two-user GMAC can be viewed as a mapping $\phi$ given by $$\phi\left(\sqrt{P_{1}} x_{1}, \sqrt{P_{2}} x_{2}\right) = \sqrt{P_{1}} x_{1} + \sqrt{P_{2}} x_{2}.$$ A constellation pair ($\mathcal{S}_{1}$, $\mathcal{S}_{2}$) is said to be uniquely decodable (UD) if the mapping $\phi$ is one-one.

\section{Constellation Power Allocation Scheme for Two-User GMAC}
\label{sec3}
In Section \ref{sec2}, we have assumed unit average power on $\mathcal{S}_{1}$ and $\mathcal{S}_{2}$. Therefore, to meet the average power constraint $P_{i}$, User-$i$ can transmit symbols of the form $\sqrt{P_{i}}x_{i}$. We now propose an alternate method to transmit the symbols of $\mathcal{S}_{i}$ by maintaining the average power constraint $P_{i}$. To explain the new scheme, we let $\mathcal{T} = \{ 1, 2, 3, \cdots, T-1, T\}$ denote the set of the indices of complex channel use, where $T$ is the total number of channel uses. Assuming $T$ to be an even number, let $\mathcal{T}_{\small{\mbox{odd}}} = \{ 1, 3, 5 \cdots, T-3, T-1\}$ and $\mathcal{T}_{\small{\mbox{even}}} = \{ 2, 4, 6 \cdots, T-2, T\}$ denote the set of odd and even indices of channel use, respectively. We use the variable $t$ to denote the instantaneous channel use index. We also use $\alpha \in [0, 1]$ to denote a real valued variable which is used to vary the transmit power of each user. Using the pseudo code representation, we explain the CPA scheme below.

\begin{figure*}
{\footnotesize
\begin{equation}
\label{thmeq}
Q(\bar{\alpha}) = \sum_{k_{1} = 0}^{N_{1} - 1}\sum_{k_{2} = 0}^{N_{2} - 1}\mbox{log}_{2}\left[ \sum_{i_{1} = 0}^{N_{1} - 1}\sum_{i_{2} = 0}^{N_{2} - 1} \mbox{exp}\left(- |\sqrt{(2-\bar{\alpha})P_{1}}(x_{1}(k_{1}) - x_{1}(i_{1})) + \sqrt{\bar{\alpha} P_{2}} ( x_{2}(k_{2}) - x_{2}(i_{2})) |^{2}/2\sigma^{2}\right)\right]
\end{equation}
$~~$\\
\begin{equation}
I^{(1)}(\sqrt{P_{L}}x_{1} + \sqrt{P_{S}}x_{2} : y) = \nonumber
\end{equation}
\begin{equation}
\label{prfeq1}
\sum_{k_{1} = 0}^{N_{1} - 1}\sum_{k_{2} = 0}^{N_{2} - 1}\underbrace{E\left[\mbox{log}_{2}\underbrace{\left[ \sum_{i_{1} = 0}^{N_{1} - 1}\sum_{i_{2} = 0}^{N_{2} - 1} \mbox{exp}\left(- |\sqrt{P_{L}}x_{1}(k_{1}) + \sqrt{P_{S}}x_{2}(k_{2}) - \sqrt{P_{L}}x_{1}(i_{1}) - \sqrt{P_{S}}x_{2}(i_{2}) + z|^{2}/\sigma^{2}\right)\right]}_{\beta(k_{1}, k_{2}, z)} \right]}_{\lambda(k_{1}, k_{2})}
\end{equation}
}
\hrule
\end{figure*}

\vspace{.2cm}
\begin{mdframed}
\noindent \textbf{IF} $t \in \mathcal{T}_{\small{\mbox{odd}}}$
\begin{itemize}
\item User-$1$ transmits $\sqrt{(2-\alpha)P_{1}}x_{1}$ for $x_{1} \in \mathcal{S}_{1}$
\item User-$2$ transmits $\sqrt{\alpha P_{2}}x_{2}$ for $x_{2} \in \mathcal{S}_{2}$
\end{itemize}
\noindent \textbf{ELSE}
\begin{itemize}
\item  User-$1$ transmits $\sqrt{\alpha P_{1}}x_{1}$ for $x_{1} \in \mathcal{S}_{1}$
\item  User-$2$ transmits $\sqrt{(2-\alpha)P_{2}}x_{2}$ for $x_{2} \in \mathcal{S}_{2}$
\end{itemize}
\noindent \textbf{END}
\end{mdframed}
\vspace{.2cm}

During odd indices, the destination receives a symbol $y_{t}$ of the form $y_{t} = \sqrt{(2-\alpha)P_{1}} x_{1} + \sqrt{\alpha P_{2}} x_{2} + z.$ Similarly, during even indices, the destination receives a symbol $y_{t}$ of the form $y_{t} = \sqrt{\alpha P_{1}} x_{1} + \sqrt{(2-\alpha) P_{2}} x_{2} + z.$ With the above mentioned power allocation, the average power for User-$1$ is $(2-\alpha)P_{1}$ and $\alpha P_{1}$ in odd and even channel uses, respectively. Similarly, the average power for User-$2$ is $\alpha P_{2}$ and $(2-\alpha)P_{2}$ in odd and even channel uses, respectively. With this, the average power $P_{i}$ is maintained for User-$i$. When the two users employ identical constellations and equal average power, an appropriate value of $\alpha$ can provide the UD property at the receiver for every channel use.


\indent We now proceed to find the optimal $\alpha$ that maximizes the CC sum capacity of the CPA scheme.

\subsection{Optimal $\alpha$ for the CPA Scheme}
\label{sec3_subsec1}
We consider the \emph{identical constellation} case, i.e., $\mathcal{S}_{1} = \mathcal{S}_{2}$, and focus on finding $\alpha$ such that the CC sum capacity is maximized. Through the CPA scheme, the two users switch the scale factors on alternate channel uses. During odd channel uses, the destination views the sum constellation $\mathcal{S}_{sum, odd}$ given by $\mathcal{S}_{sum, odd} \triangleq \left\lbrace \sqrt{(2-\alpha)P_{1}}x_{1} + \sqrt{\alpha P_{2}}x_{2} ~ | ~ \forall x_{1},  x_{2}\right\rbrace.$ During even channel uses, the destination views $\mathcal{S}_{sum, even}$ given by $\mathcal{S}_{sum, even} \triangleq \left\lbrace \sqrt{\alpha P_{1}}x_{1} + \sqrt{(2-\alpha) P_{2}}x_{2} ~ | ~ \forall x_{1},  x_{2}\right\rbrace.$ If the two users have equal average power constraint, the destination views the same sum constellation on every channel use. On the other hand, if the two users have unequal average power constraint, then the sum constellation seen by the destination is \emph{not the same} during the even and odd indices of the channel use. Considering the general case of unequal average power constraints, the CC sum capacity of the CPA scheme is given by
\begin{equation*}
\frac{1}{2} \sum_{\bar{\alpha} \in \Omega} I\left(\sqrt{(2-\bar{\alpha})P_{1}}x_{1} + \sqrt{\bar{\alpha} P_{2}}x_{2} : y\right),
\end{equation*}
where $\Omega = \{ \alpha, 2-\alpha \}$. Note that the CC sum capacity can be increased by choosing an appropriate $\alpha \in [0, 1]$. However, we know that $\alpha = 0$ corresponds to single-user transmission. As a result, henceforth, we consider selecting an appropriate $\alpha$ in the interval $(0, 1]$. Thus, the CC sum capacity can be maximized by choosing $\alpha_{\small{\mbox{opt}}}$ given by,
\begin{equation}
\label{max_problem}
\alpha_{\small{\mbox{opt}}} = \argmax_{\alpha \in (0, 1]} \frac{1}{2} \sum_{\bar{\alpha} \in \Omega} I\left(\sqrt{(2-\bar{\alpha})P_{1}}x_{1} + \sqrt{\bar{\alpha} P_{2}}x_{2} : y\right).
\end{equation}
Note that the above objective function involves expectation of a non-linear function of the random variable $z$, and hence, its closed form expression is not available. Therefore, computing $\alpha_{\small{\mbox{opt}}}$ is not straightforward. For high values of $\frac{P_{1}}{\sigma^2}$ and $\frac{P_{2}}{\sigma^2}$, the following theorem provides a deterministic metric (which is independent of the variable $z$) to choose $\alpha$ such that the CC capacity is maximized. Henceforth, high SNR values imply high values of $\frac{P_1}{\sigma^2}$ and $\frac{P_2}{\sigma^2}$.

\begin{theorem}
\label{thm}
At high SNR values, the optimum scale factor $\alpha$ required to maximize the CC sum capacity is approximated closely by $\alpha^{*}$ where
\begin{equation*}
\alpha^{*} = \argmin_{\alpha \in (0, 1]} \sum_{\bar{\alpha} \in \Omega} Q(\bar{\alpha}),
\end{equation*}
where $Q(\bar{\alpha})$ is given in \eqref{thmeq} at the top of this page.
\end{theorem}

\begin{proof}
Since $N_{1}$ and $N_{2}$ are constants, we have the following equality, 
\begin{align*}
&\argmax_{\alpha \in (0, 1]} \sum_{\bar{\alpha} \in \Omega} I(\sqrt{(2-\bar{\alpha})P_{1}}x_{1} + \sqrt{\bar{\alpha} P_{2}}x_{2} : y)\\ 
&= \argmin_{\alpha \in (0, 1]} \sum_{\bar{\alpha} \in \Omega} I^{(1)}(\sqrt{(2-\bar{\alpha})P_{1}}x_{1} + \sqrt{\bar{\alpha} P_{2}}x_{2} : y),
\end{align*}
where $I^{(1)}(\sqrt{(2-\bar{\alpha})P_{1}}x_{1} + \sqrt{\bar{\alpha} P_{2}}x_{2} : y)$ is given in \eqref{prfeq1}, and we use $P_{L}$ to denote $(2-\bar{\alpha})P_{1}$, and $P_{S}$ to denote $\bar{\alpha} P_{2}$. Note that the individual terms $\lambda(k_{1}, k_{2})$ of $I^{(1)}(\sqrt{P_{L}}x_{1} + \sqrt{P_{S}}x_{2}: y)$ are of the form $E\left[\mbox{log}_{2}(\beta(k_{1}, k_{2}, z))\right]$ for a random variable $\beta(k_{1}, k_{2}, z)$. Applying Jensen's inequality: $$E[\mbox{log}_{2}(\beta(k_{1}, k_{2}, z))] \leq \mbox{log}_{2}[E(\beta(k_{1}, k_{2}, z))]$$ on $\lambda(k_{1}, k_{2})$, we have
\begin{equation}
\label{theta_upper_bound}
I^{(1)}(\sqrt{P_{L}}x_{1} + \sqrt{P_{S}}x_{2}: y)  \leq Q(\bar{\alpha}),
\end{equation}
where $Q(\bar{\alpha})$ is given by \eqref{thmeq}. In the rest of the proof, we show that at high SNR values, the following approximation holds good:
\begin{align*}
\argmin_{\alpha \in (0, 1]}  \sum_{\bar{\alpha} \in \Omega} Q(\bar{\alpha}) \approx \argmin_{\alpha \in (0, 1]} \sum_{\bar{\alpha} \in \Omega} I^{(1)}(\sqrt{P_{L}}x_{1} + \sqrt{P_{S}}x_{2}: y).
\end{align*}

\begin{figure*}
\footnotesize{
\begin{equation}
\label{proof_num1}
I^{(1)}(\sqrt{P_{L}}x_{1} + \sqrt{P_{S}}x_{2} : y) = \sum_{k_{1} = 0}^{N_{1} - 1}\sum_{k_{2} = 0}^{N_{2} - 1}E\left[\mbox{log}_{2}\left[\mbox{exp}\left(- |z|^{2}/\sigma^{2}\right) + \underbrace{\sum_{i_{1} = 0}^{N_{1} - 1}\sum_{i_{2} = 0}^{N_{2} - 1}}_{(i_{1}, i_{2}) \neq (k_{1}, k_{2})} \mbox{exp}\left(- |\mu(k_{1}, k_{2}, i_{1}, i_{2}) + z|^{2}/\sigma^{2}\right)\right]\right]
\end{equation}
\begin{equation}
\label{proof_num2}
 = \sum_{k_{1} = 0}^{N_{1} - 1}\sum_{k_{2} = 0}^{N_{2} - 1}E\left[\mbox{log}_{2}\left[\mbox{exp}\left(- |z|^{2}/\sigma^{2}\right)\left(1 + M(k_{1}, k_{2})\right)+ \underbrace{\sum_{i_{1} = 0}^{N_{1} - 1}\sum_{i_{2} = 0}^{N_{2} - 1}}_{\mu(k_{1}, k_{2}, i_{1}, i_{2}) \neq 0 } \mbox{exp}\left(- |\mu(k_{1}, k_{2}, i_{1}, i_{2}) + z|^{2}/\sigma^{2}\right)\right]\right]
\end{equation}
}
\end{figure*}

\noindent Note that the term $I^{(1)}(\sqrt{P_{L}}x_{1} + \sqrt{P_{S}}x_{2}: y)$ can be written as in \eqref{proof_num1} and \eqref{proof_num2} where
\begin{align*}
\mu(k_{1}, k_{2}, i_{1}, i_{2}) = & \sqrt{P_{L}}x_{1}(k_{1}) + \sqrt{P_{S}}x_{2}(k_{2})\\
&  - \sqrt{P_{L}}x_{1}(i_{1}) - \sqrt{P_{S}}x_{2}(i_{2}),
\end{align*}
\noindent and $M(k_{1}, k_{2}) = |\mathcal{M}(k_{1}, k_{2})|$ such that $\mathcal{M}(k_{1}, k_{2})$ is given by
\begin{equation*}
\mathcal{M}(k_{1}, k_{2}) = \{ (i_{1}, i_{2}) \neq (k_{1}, k_{2}) ~|~ \mu(i_{1}, i_{2}, k_{1}, k_{2}) = 0\}.
\end{equation*}
\noindent Removing independent terms of the form $\mbox{exp}\left(- \frac{|z|^{2}}{\sigma^{2}}\right)$ in \eqref{proof_num2}, we have
\begin{align*}
& \argmin_{\alpha \in (0, 1]} \sum_{\bar{\alpha} \in \Omega} I^{(1)}(\sqrt{P_{L}}x_{1} + \sqrt{P_{S}}x_{2}: y)\\
& = \argmin_{\alpha \in (0, 1]} \sum_{\bar{\alpha} \in \Omega} I^{(2)}(\sqrt{P_{L}}x_{1} + \sqrt{P_{S}}x_{2}: y),
\end{align*}
where $I^{(2)}(\sqrt{P_{L}}x_{1} + \sqrt{P_{S}}x_{2}: y)$ is given in \eqref{proof_num3}. Further, at high SNR values, we have the approximation,
\begin{align*}
& \argmin_{\alpha \in (0, 1]} \sum_{\bar{\alpha} \in \Omega} I^{(2)}(\sqrt{P_{L}}x_{1} + \sqrt{P_{S}}x_{2}: y)\\ & \approx \argmin_{\alpha \in (0, 1]} \sum_{\bar{\alpha} \in \Omega} I^{(3)}(\sqrt{P_{L}}x_{1} + \sqrt{P_{S}}x_{2}: y),
\end{align*}
where $I^{(3)}(\sqrt{P_{L}}x_{1} + \sqrt{P_{S}}x_{2}: y)$ is given by \eqref{proof_num4}. At high SNR values, each term $\gamma(k_{1}, k_{2}, z)$ in \eqref{proof_num4} is small, and hence we use the approximation $\mbox{log}_{2}(1 + \gamma(k_{1}, k_{2}, z)) \approx \mbox{log}_{2}(e)(\gamma(k_{1}, k_{2}, z))$ to obtain \eqref{proof_num5}. Evaluating the expectation in \eqref{proof_num5}, we get \eqref{proof_num6}. Once again, applying the approximation $\mbox{log}_{2}(1 + \delta(k_{1}, k_{2})) \approx \mbox{log}_{2}(e)(\delta(k_{1}, k_{2}))$ in \eqref{proof_num6}, we get \eqref{proof_num7}, which is denoted by $I^{(4)}(\sqrt{P_{L}}x_{1} + \sqrt{P_{S}}x_{2}: y)$.\\
\indent Now, we consider the term $I^{(5)}(\sqrt{P_{L}}x_{1} + \sqrt{P_{S}}x_{2}: y)$ given in \eqref{proof_num8} and show the following equality:
\begin{align}
\label{critical_equality}
& \argmin_{\alpha \in (0, 1]} \sum_{\bar{\alpha} \in \Omega} I^{(5)}(\sqrt{P_{L}}x_{1} + \sqrt{P_{S}}x_{2}: y) \nonumber \\ & = \argmin_{\alpha \in (0, 1]} \sum_{\bar{\alpha} \in \Omega} I^{(4)}(\sqrt{P_{L}}x_{1} + \sqrt{P_{S}}x_{2}: y).
\end{align}
Once the above equality is proved, the statement of this theorem also gets proved since $I^{(5)}(\sqrt{P_{L}}x_{1} + \sqrt{P_{S}}x_{2}: y)$ is a scaled version of $Q(\bar{\alpha})$ (as shown in \eqref{proof_num9}). 
Towards proving the equality in \eqref{critical_equality}, note that at high SNR values, $\delta(k_{1}, k_{2})$ is small for all values of $\alpha$.
\begin{figure*}
\footnotesize{
\begin{equation*}
I^{(2)}(\sqrt{P_{L}}x_{1} + \sqrt{P_{S}}x_{2} : y) =
\end{equation*}
\begin{equation}
\label{proof_num3}
\sum_{k_{1} = 0}^{N_{1} - 1}\sum_{k_{2} = 0}^{N_{2} - 1}E\left[\mbox{log}_{2}\left(1 + M(k_{1}, k_{2})\right) + \mbox{log}_{2}\left[1 + \frac{1}{\left(1 + M(k_{1}, k_{2})\right)}\underbrace{\sum_{i_{1} = 0}^{N_{1} - 1}\sum_{i_{2} = 0}^{N_{2} - 1}}_{\mu(k_{1}, k_{2}, i_{1}, i_{2}) \neq 0 } \mbox{exp}\left( \frac{-|\mu(k_{1}, k_{2}, i_{1}, i_{2}) + z|^{2} + |z|^{2}}{\sigma^{2}}\right)\right]\right]
\end{equation}
\begin{equation*}
 I^{(3)}(\sqrt{P_{L}}x_{1} + \sqrt{P_{S}}x_{2} : y) =
\end{equation*}
\begin{equation}
\label{proof_num4}
\sum_{k_{1} = 0}^{N_{1} - 1}\sum_{k_{2} = 0}^{N_{2} - 1}E\left[\mbox{log}_{2}\left(1 + M(k_{1}, k_{2})\right) + \mbox{log}_{2}\left[1 + \underbrace{\frac{1}{\left(1 + M(k_{1}, k_{2})\right)}\underbrace{\sum_{i_{1} = 0}^{N_{1} - 1}\sum_{i_{2} = 0}^{N_{2} - 1}}_{\mu(k_{1}, k_{2}, i_{1}, i_{2}) \neq 0 } \mbox{exp}\left( \frac{-|\mu(k_{1}, k_{2}, i_{1}, i_{2}) + z|^{2}}{\sigma^{2}}\right)}_{\gamma(k_{1}, k_{2}, z)}\right]\right]
\end{equation}
}
\hrule
\end{figure*}
For those values of $\alpha$ which provide the UD property, we have $\mbox{log}_{2}(1 + M(k_{1}, k_{2})) = 0 ~\forall k_{1}, k_{2}$. However, for those values of $\alpha$ which \emph{do not} provide the UD property, we have $\mbox{log}_{2}(1 + M(k_{1}, k_{2})) \geq 1$ for some $k_{1}, k_{2}$. Further, at high SNR, for all $\alpha$, $\mbox{log}_{2}(1+\delta(k_1,k_2))<<1,~ \forall k_1,k_2$. Due to these reasons, the values of $\alpha$ which do not provide the UD property do not minimize $I^{(5)}(\sqrt{P_{L}}x_{1} + \sqrt{P_{S}}x_{2}: y)$ as well as $I^{(4)}(\sqrt{P_{L}}x_{1} + \sqrt{P_{S}}x_{2}: y)$. As a result, the optimal value of $\alpha$ must belong to the set which provides the UD property. For such values of $\alpha$, we have $I^{(5)}(\sqrt{P_{L}}x_{1} + \sqrt{P_{S}}x_{2}: y) = I^{(4)}(\sqrt{P_{L}}x_{1} + \sqrt{P_{S}}x_{2}: y),$ and hence the equality in \eqref{critical_equality} holds. This completes the proof.
\end{proof}

\begin{figure*}
{\footnotesize
\begin{equation*}
 I^{(3)}(\sqrt{P_{L}}x_{1} + \sqrt{P_{S}}x_{2} : y) \approx
\end{equation*}
\begin{equation}
\label{proof_num5}
\sum_{k_{1} = 0}^{N_{1} - 1}\sum_{k_{2} = 0}^{N_{2} - 1} \left[\mbox{log}_{2}\left(1 + M(k_{1}, k_{2})\right) + E\left[\frac{\mbox{log}_{2}(e)}{\left(1 + M(k_{1}, k_{2})\right)}\underbrace{\sum_{i_{1} = 0}^{N_{1} - 1}\sum_{i_{2} = 0}^{N_{2} - 1}}_{\mu(k_{1}, k_{2}, i_{1}, i_{2}) \neq 0 } \mbox{exp}\left( \frac{-|\mu(k_{1}, k_{2}, i_{1}, i_{2}) + z|^{2}}{\sigma^{2}}\right)\right]\right]
\end{equation}
\begin{equation}
\label{proof_num6}
= \sum_{k_{1} = 0}^{N_{1} - 1}\sum_{k_{2} = 0}^{N_{2} - 1}\left[\mbox{log}_{2}\left(1 + M(k_{1}, k_{2})\right) + \frac{\mbox{log}_{2}(e)}{2} \left[\underbrace{\frac{1}{\left(1 + M(k_{1}, k_{2})\right)}\underbrace{\sum_{i_{1} = 0}^{N_{1} - 1}\sum_{i_{2} = 0}^{N_{2} - 1}}_{\mu(k_{1}, k_{2}, i_{1}, i_{2}) \neq 0 } \mbox{exp}\left( \frac{-|\mu(k_{1}, k_{2}, i_{1}, i_{2})|^{2}}{2\sigma^{2}}\right)}_{\delta(k_{1}, k_{2})}\right]\right]
\end{equation}
\begin{equation*}
\label{proof_num7}
I^{(4)}(\sqrt{P_{L}}x_{1} + \sqrt{P_{S}}x_{2} : y) =
\end{equation*}
\begin{equation}
\label{proof_num7}
\sum_{k_{1} = 0}^{N_{1} - 1}\sum_{k_{2} = 0}^{N_{2} - 1}\left[\mbox{log}_{2}\left(1 + M(k_{1}, k_{2})\right) + \frac{1}{2}\mbox{log}_{2}\left[1 + \frac{1}{\left(1 + M(k_{1}, k_{2})\right)}\underbrace{\sum_{i_{1} = 0}^{N_{1} - 1}\sum_{i_{2} = 0}^{N_{2} - 1}}_{\mu(k_{1}, k_{2}, i_{1}, i_{2}) \neq 0 } \mbox{exp}\left( \frac{-|\mu(k_{1}, k_{2}, i_{1}, i_{2})|^{2}}{2\sigma^{2}}\right)\right]\right]
\end{equation}
\begin{equation*}
I^{(5)}(\sqrt{P_{L}}x_{1} + \sqrt{P_{S}}x_{2} : y) =
\end{equation*}
\begin{equation}
\label{proof_num8}
\sum_{k_{1} = 0}^{N_{1} - 1}\sum_{k_{2} = 0}^{N_{2} - 1}\frac{1}{2}\left[\mbox{log}_{2}\left(1 + M(k_{1}, k_{2})\right) + \mbox{log}_{2}\left[1 + \frac{1}{\left(1 + M(k_{1}, k_{2})\right)}\underbrace{\sum_{i_{1} = 0}^{N_{1} - 1}\sum_{i_{2} = 0}^{N_{2} - 1}}_{\mu(k_{1}, k_{2}, i_{1}, i_{2}) \neq 0 } \mbox{exp}\left( \frac{-|\mu(k_{1}, k_{2}, i_{1}, i_{2})|^{2}}{2\sigma^{2}}\right)\right]\right]
\end{equation}
\begin{equation}
\label{proof_num9}
= \frac{1}{2}\sum_{k_{1} = 0}^{N_{1} - 1}\sum_{k_{2} = 0}^{N_{2} - 1}\mbox{log}_{2}\left[ \sum_{i_{1} = 0}^{N_{1} - 1}\sum_{i_{2} = 0}^{N_{2} - 1} \mbox{exp}\left(- |\sqrt{P_{L}}x_{1}(k_{1}) - \sqrt{P_{L}}x_{1}(i_{1}) + (\sqrt{P_{S}}x_{2}(k_{2})  - \sqrt{P_{S}}x_{2}(i_{2})) |^{2}/2\sigma^{2}\right)\right] = \frac{1}{2}Q(\bar{\alpha})
\end{equation}
}
\hrule
\end{figure*}

\indent Using the results of Theorem \ref{thm}, for high SNR values, we propose to find $\alpha^{*}$ which minimizes $\sum_{\bar{\alpha} \in \Omega} Q(\bar{\alpha})$, a tight upper bound on $\sum_{\bar{\alpha} \in \Omega} I^{(1)}(\sqrt{P_{L}}x_{1} + \sqrt{P_{S}}x_{2}: y)$.
However, note that for small to moderate values of SNR, the values of $\alpha^{*}$ obtained by solving \eqref{thmeq} need not maximize $\sum_{\bar{\alpha} \in \Omega} I(\sqrt{P_{L}}x_{1} + \sqrt{P_{S}}x_{2}: y)$ since the bound in \eqref{theta_upper_bound} is not known to be tight. It is also clear that solving \eqref{thmeq} is easier than solving \eqref{max_problem} since $\sum_{\bar{\alpha} \in \Omega} Q(\bar{\alpha})$ is deterministic and independent of the term $z$. 

\subsection{Numerical Results}
\label{sec3_subsec2}

\begin{figure}[h]
\centering
\includegraphics[width=3in]{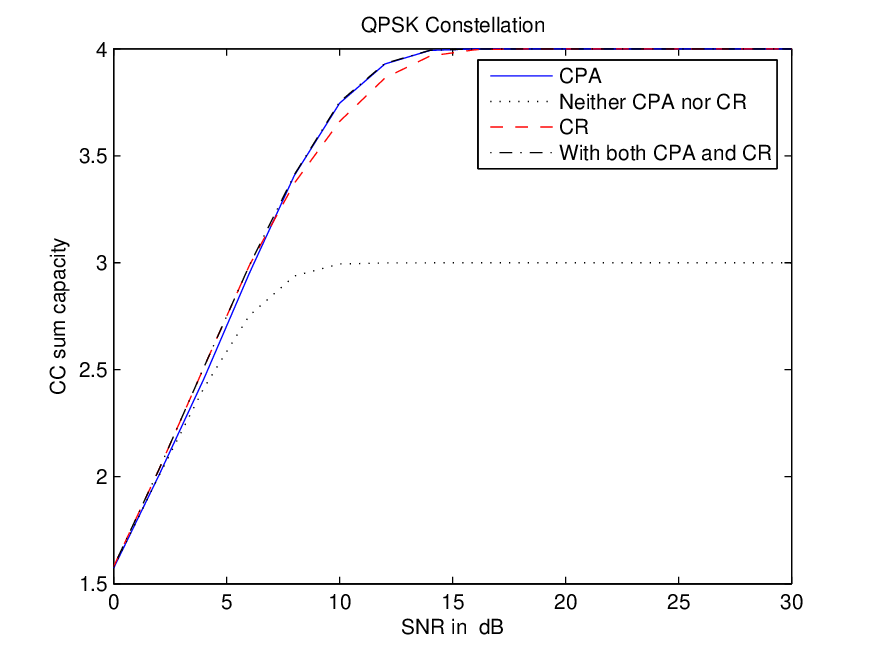}
\caption{CC sum capacity of QPSK constellation with equal average power constraint and no random phase offsets.}
\label{plot1}
\end{figure}

\begin{figure}[h]
\centering
\includegraphics[width=3in]{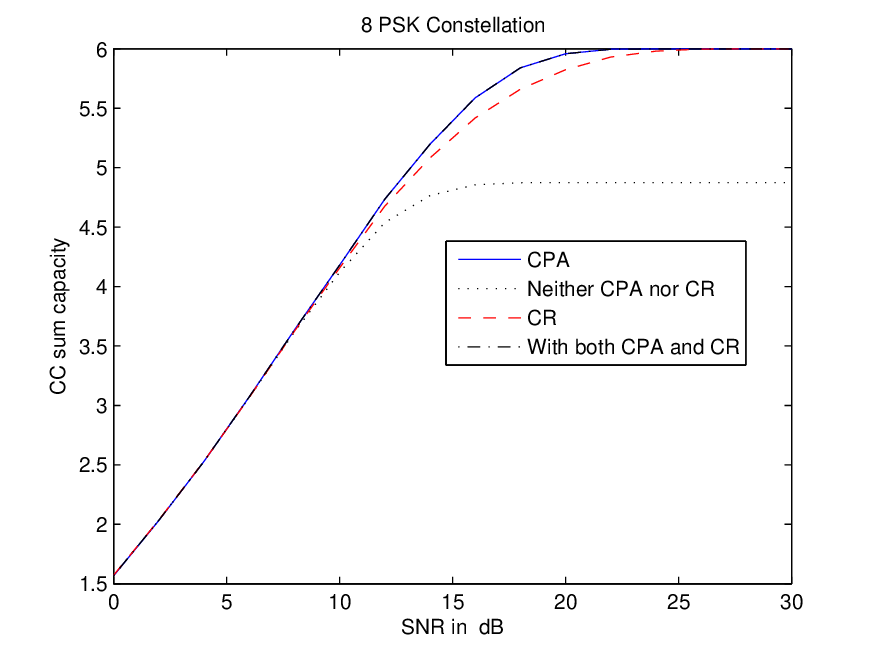}
\caption{CC sum capacity of 8-PSK constellation with equal average power constraint and no random phase offsets.}
\label{plot2}
\end{figure}

In this section, we first compute $\alpha^{*}$ for the case of equal average power constraint for the two users (i.e., $P_{1} = P_{2}$). For the simulation results, we use $\sigma^{2} = 1$ and $\mbox{SNR} = P_{1}$. The values of $\alpha^{*}$ are obtained by varying $\alpha$ from 0 to 1 in steps of 0.01. In Table \ref{table_for_scale_gmac}, the values of $\alpha^{*}$ are presented for some known constellations. The CC sum capacities of QPSK and $8$-PSK are also provided in Fig. \ref{plot1} and Fig. \ref{plot2}, respectively for the following schemes: ({\em i}) CPA scheme, ({\em ii}) CR scheme, ({\em iii}) neither CPA nor CR, and ({\em iv}) with both CPA and CR. For the scheme ``with both CPA and CR", the pairs ($\alpha^{*}$, $\theta^{*}$) are computed using a metric which is obtained on the similar lines of Theorem \ref{thm}. From the figures, note that the CPA scheme provides CC sum capacities close (or equal) to the CR scheme for all SNR values. Note that the scheme ``with both CPA and CR" does not provide significant CC capacity gain over the CPA scheme. 

\begin{figure}[h]
\centering
\includegraphics[width=3in]{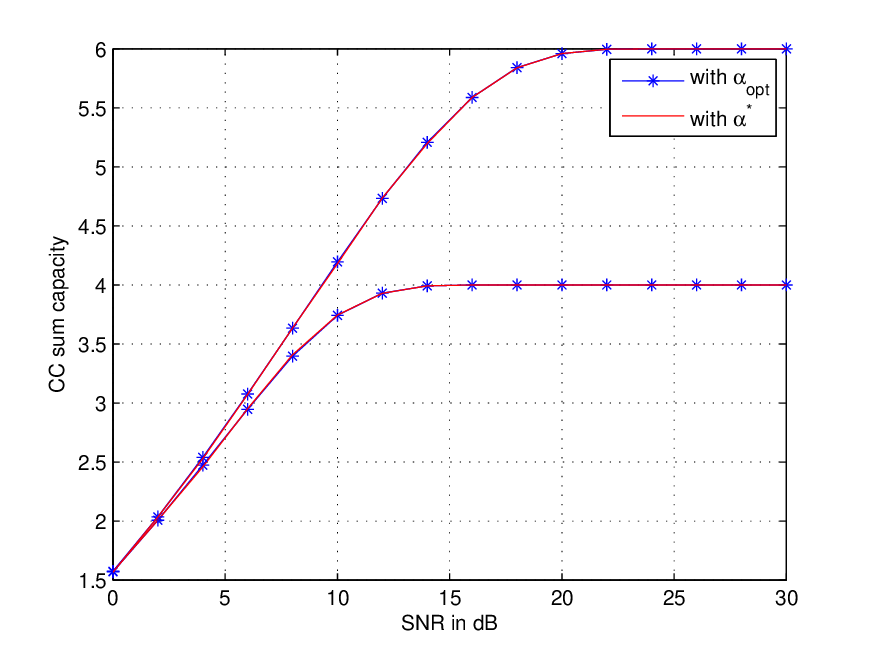}
\caption{CC sum capacity of QPSK and 8-PSK constellations with $\alpha_{\small{\mbox{opt}}}$ and $\alpha^{*}$ for GMAC with equal average power constraint and no random phase offsets.}
\label{metric_goodness}
\end{figure}

\indent To verify the results of Theorem \ref{thm}, we plot the CC sum capacity of QPSK and 8-PSK with $\alpha_{\small{\mbox{opt}}}$ and $\alpha^{*}$ in Fig. \ref{metric_goodness}. From the figure, we see that $\alpha^{*}$ provides approximately same CC sum capacity values not only at high SNR but also at moderate SNR values. In addition, we also present the corresponding values of $\alpha_{\small{\mbox{opt}}}$ and $\alpha^{*}$ in Table. \ref{values_of_alphas}, which shows that $\alpha_{\small{\mbox{opt}}}$ and $\alpha^{*}$ are different for low SNR values, and are approximately same for high SNR values. These results demonstrate that the metric $\sum_{\bar{\alpha} \in \Omega} Q(\bar{\alpha})$ is a tight upper bound on $\sum_{\bar{\alpha} \in \Omega} I^{(1)}(\sqrt{P_{L}}x_{1} + \sqrt{P_{S}}x_{2}: y)$ at high SNR values.

\begin{figure}[h]
\centering
\includegraphics[width=3in]{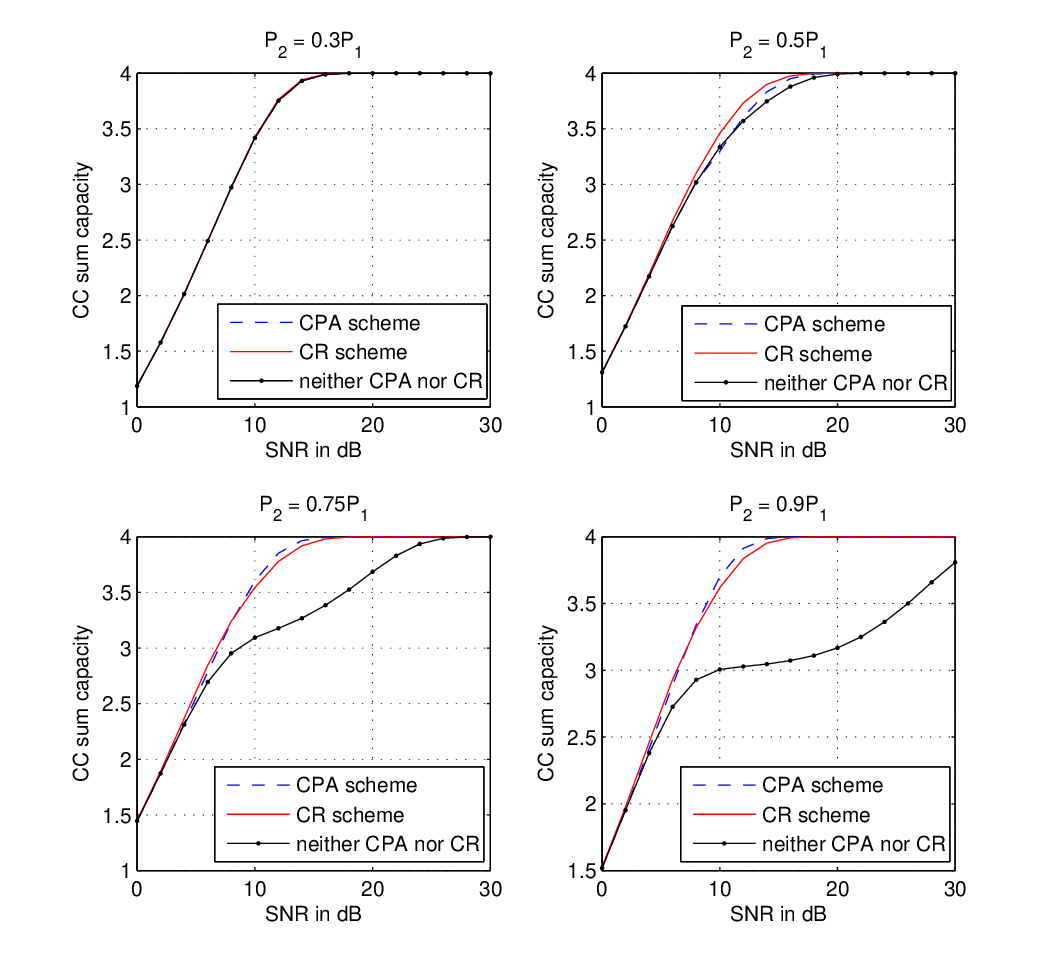}
\caption{CC sum capacities of QPSK with unequal average power constraint and no random phase offsets.}
\label{uneqp_np}
\end{figure}

\indent For the unequal average power case, we compute $\alpha^{*}$ for the QPSK constellation. We use $\mbox{SNR}  = P_{1}$ (with $\sigma^{2} = 1$). For the simulation results, we have considered the following relations between $P_{1}$ and $P_{2}$: ({\em i}) $P_{2}$ = $0.3P_{1}$, ({\em ii}) $P_{2}$ = $0.5P_{1}$ and ({\em iii}) $P_{2}$ = $0.75P_{1}$, and ({\em iv}) $P_{2}$ = $0.9P_{1}$. In Fig. \ref{uneqp_np}, the CC sum capacities of QPSK are provided for the CPA scheme, the CR scheme, and the ``neither CPA nor CR" scheme. When compared to the ``neither CPA nor CR" scheme, the CPA scheme provides increased CC sum capacities at moderate to high SNR values, when $P_{1}$ and $P_{2}$ are close. When compared to the CR scheme, the CPA scheme provides marginal improvement in the CC sum capacities only at high SNR values, when $P_{1}$ and $P_{2}$ are close. The figure shows that the gains of the CPA scheme diminishes when the power difference is more than $3$ dB (as shown for the case $P_{2}$ = $0.5P_{1}$ and $P_{2}$ = $0.3P_{1}$).

\begin{center}
\begin{table*}
\caption{Numerically computed $\alpha^{*}$ with equal average power constraint and no random phase offsets.}
\begin{center}
\begin{tabular}{|c|c|c|c|c|c|c|c|c|c|c|}
\hline SNR in dB & QPSK & 8-PSK & 16-PSK & 16-QAM\\
\hline 0 & 0.74 & 0.65 & 0.65 & 0.48\\
\hline 2 & 0.65 & 0.85 & 0.85 & 1.00\\
\hline 4 & 0.52 & 0.74 & 0.74 & 1.00\\
\hline 6 & 0.46 & 0.67 & 0.67 & 1.00\\
\hline 8 & 0.43 & 0.65 & 0.66 & 0.84\\
\hline 10 & 0.41 & 0.64 & 0.67 & 0.74\\
\hline 12 & 0.41 & 0.59 & 0.70 & 0.68\\
\hline 14 & 0.40 & 0.53 & 0.74 & 0.46\\
\hline 16 & 0.40 & 0.50 & 0.79 & 0.62\\
\hline 18 & 0.40 & 0.49 & 0.78 & 0.13\\
\hline 20 & 0.37 & 0.49 & 0.59 & 0.12\\
\hline 22 & 0.24 & 0.49 & 0.58 & 0.12\\
\hline 24 & 0.15 & 0.49 & 0.56 & 0.12\\
\hline 26 & 0.10 & 0.49 & 0.55 & 0.12\\
\hline 28 & 0.06 & 0.48 & 0.55 & 0.12\\
\hline 30 & 0.04 & 0.13 & 0.55 & 0.12\\
\hline
\end{tabular}
\end{center}
\label{table_for_scale_gmac}
\end{table*}
\end{center}

\begin{center}
\begin{table*}
\caption{Numerically computed values of $\alpha$ with equal average power constraint and no random phase offsets.}
\begin{center}
\begin{tabular}{|c|c|c|c|c|c|c|c|c|c|c|}
\hline SNR in dB & $\alpha_{\small{\mbox{opt}}}$ for QPSK & $\alpha^{*}$ for QPSK & $\alpha_{\small{\mbox{opt}}}$ for 8-PSK & $\alpha^{*}$ for 8-PSK \\
\hline 0 & 0.88 & 0.74 & 0.92 & 0.65\\
\hline 4 & 0.55 & 0.52 & 0.83 & 0.74\\
\hline 8 & 0.44 & 0.43 &  0.66 & 0.65\\
\hline 12 & 0.41 & 0.41 & 0.60 & 0.59\\
\hline 16 & 0.41 & 0.40 & 0.52 & 0.50\\
\hline 20 & 0.38  & 0.37 & 0.49 & 0.49\\
\hline 24 & 0.19  & 0.15 & 0.49 & 0.49\\
\hline
\end{tabular}
\end{center}	 
\label{values_of_alphas}
\end{table*}
\end{center}

\subsection{Reduced ML Decoding Complexity for QAM Constellations with CPA}
\label{sec3_subsec4}

%

In this subsection, we highlight the advantage of using CPA over CR for the class of regular QAM constellations. For uncoded transmission, when CR is employed for QAM constellations, the in-phase and the quadrature components of the symbols of $\mathcal{S}_{sum}$ are entangled. However, with CPA, since the scale factor $\alpha$ is real valued, the in-phase and quadrature components of the symbols in $\mathcal{S}_{sum}$ \emph{do not} get entangled. As a result, $\mathcal{S}_{sum}$ is separable, and can be written as the cross product of in-phase and quadrature components of its symbols. In particular, if the two users employ square regular $M$-QAM constellation, then there are $M$ points in $\mathcal{S}_{sum}$ along the in-phase and the quadrature component, respectively. Since $\mathcal{S}_{sum}$ is separable, the destination can decode the in-phase and the quadrature components independently. Therefore, the worst case ML decoding complexity is $O(M)$. However, with CR scheme, the worst case ML decoding complexity is $O(M^{2})$. Therefore, for the class of QAM constellations, CPA provides lower decoding complexity with negligible loss in the CC sum capacity when compared with CR.

\section{Channel Coding for QAM Constellations with CPA}
\label{sec4}
In this section, we design code pairs based on TCM (Trellis Coded Modulation) \cite{Ub} to achieve sum-rates close to the CC sum-capacity of QAM constellations using the CPA scheme. Throughout this section, we assume equal average power constraints for the two users with $P_{1} = P_{2} = P$. If $\mathcal{S}_{1}$ and $\mathcal{S}_{2}$ represent regular $M$-QAM constellations, then the in-phase and the quadrature components of $\mathcal{S}_{sum}$ are respectively of the form
$\sqrt{(2-\alpha)P} x_{1I} + \sqrt{\alpha P} x_{2I}$ and $\sqrt{(2-\alpha)P} x_{1Q} + \sqrt{\alpha P} x_{2Q}$, where $x_{iI} \in \mathcal{S}_{iI}, x_{iQ} \in \mathcal{S}_{iQ}$ and $\mathcal{S}_{iI}$ and $\mathcal{S}_{iQ}$ denote the corresponding $\sqrt{M}$-PAM constellations for the $i$-th user along the in-phase and the quadrature dimension, respectively. The set of in-phase and quadrature symbols of $\mathcal{S}_{sum}$ are respectively denoted by $\mathcal{S}_{sum,I}$ and $\mathcal{S}_{sum,Q}$, and are given by $\mathcal{S}_{sum,I} = \left\lbrace (\sqrt{(2-\alpha)P} x_{1I} + \sqrt{\alpha P} x_{2I}) ~|~ x_{iI} \in \mathcal{S}_{iI} \right\rbrace,$ and $\mathcal{S}_{sum,Q} = \left\lbrace (\sqrt{(2-\alpha)P} x_{1Q} + \sqrt{\alpha P} x_{2Q}) ~|~ x_{iQ} \in \mathcal{S}_{iQ} \right\rbrace.$ Since the CPA scheme makes the in-phase and quadrature components of $\mathcal{S}_{sum}$ separable, the symbols of $\mathcal{S}_{sum,I}$ can be decoded independent of the symbols of $\mathcal{S}_{sum,Q}$, thereby reducing the decoding complexity. To facilitate this, User-$i$ should make sure that the symbols of $\mathcal{S}_{iQ}$ and $\mathcal{S}_{iI}$ are not coded jointly. Therefore, each user can have two encoders one along each dimension. Let the subscript $\small{X} \in \{\small{I}, \small{Q}\}$ denote either the in-phase dimension or quadrature dimension. For each $i \in \{1, 2\}$, let User-$i$ be equipped with a convolutional encoder $C_{i\small{X}}$ with $m_{iX}$ input bits and $m_{iX} + 1$ output bits. Throughout the section, we consider convolutional codes which add only 1-bit redundancy. Let the $m_{iX} + 1$ output bits of $C_{iX}$ take values from $\sqrt{M}$-PAM constellation $\mathcal{S}_{iX}$ such that $|\mathcal{S}_{iX}| = 2^{m_{iX} + 1}$. Henceforth, the set of codewords generated from $C_{1X}$ and $C_{2X}$ are represented by trellises $T_{1X}$ and $T_{2X}$ respectively.

\indent We assume that the destination performs joint decoding of the in-phase symbols of User-$1$ and User-$2$ by decoding for a sequence over $\mathcal{S}_{sum,I}$ on the sum trellis, $T_{sum,I}$ (see \cite{HaR_TIT} for the definition of the sum trellis). Similarly, joint decoding of the quadrature symbols of User-$1$ and User-$2$ is performed by decoding for a sequence over $\mathcal{S}_{sum,Q}$ on the sum trellis, $T_{sum,Q}$. Due to the existence of an equivalent AWGN channel in the GMAC set-up, the sum trellis, $T_{sum,X}$ has to be labeled with the elements of $\mathcal{S}_{sum, X}$ satisfying the design rules in \cite{Ub}. However, such a labeling rule can be obtained on $T_{sum, X}$ only through the pairs $(T_{1X}, T_{2X})$ and $(\mathcal{S}_{1X}, \mathcal{S}_{2X})$. Hence, we propose labeling rules on $T_{1X}$ and $T_{2X}$ using $\mathcal{S}_{1X}$ and $\mathcal{S}_{2X}$ respectively such that $T_{sum,X}$ is labeled with the elements of $\mathcal{S}_{sum,X}$ as per Ungerboeck rules.

\indent Since the number of input bits to $C_{iX}$ is $m_{iX}$, there are $2^{m_{iX}}$ edges diverging from (or converging to; henceforth, we only refer to diverging edges) each state of $T_{iX}$. Also, as there is only one bit redundancy added by the encoder, and as $|\mathcal{S}_{iX}| = 2^{m_{iX} + 1}$, the edges diverging from each state have to be labeled with the elements of a subset of $\mathcal{S}_{iX}$ of size $2^{m_{iX}}$. Therefore, for each $i$, $\mathcal{S}_{iX}$ has to be partitioned into two sets $\mathcal{S}^{1}_{iX}$ and $\mathcal{S}^{2}_{iX}$, and the diverging edges from each state of $T_{iX}$ have to be labeled with the elements of either $\mathcal{S}^{1}_{iX}$ or $\mathcal{S}^{2}_{iX}$. From the definition of sum trellis, there are $2^{m_{1X} + m_{2X}}$ edges diverging from each state of $T_{sum,X}$ and these edges get labeled with the elements of one of the following sets, $\mathcal{A} = \left\lbrace \mathcal{S}^{i}_{1X} + \mathcal{S}^{j}_{2X} ~|~ i, j \in \{1, 2\} \right\rbrace.$ To satisfy Ungerboeck design rules, the transitions originating from the same state of $T_{sum,X}$ must be assigned symbols that are separated by largest minimum distance. Therefore, the problem addressed is to find a partitioning of $\mathcal{S}_{iX}$ ($\sqrt{M}$-PAM constellations) into two sets $\mathcal{S}^{1}_{iX}$ and  $\mathcal{S}^{2}_{iX}$ of equal cardinality such that the minimum Euclidean distance (denoted by $d_{min}$) of each one of the sets in $\mathcal{A}$ is maximized. However, since $d_{min}$ values of the sets in $\mathcal{A}$ can potentially be different, we find a partitioning such that the minimum of the $d_{min}$ values of the sets in $\mathcal{A}$ is maximized.

\subsection{Designing TCM Schemes with 16-QAM Constellation}
\label{sec4_subsec1}
The \emph{set partitioning problem} described above depends on the structure of the sum constellation. As a result, the solution to the \emph{set partitioning problem} depends on the choice of $\alpha$. For arbitrary values of $M$ and $\alpha$, we are unable to solve the set partitioning problem due to lack of structure on $\mathcal{S}_{sum}$ of two QAM constellations. However, through computer simulations, we have found a solution for the above problem for 16-QAM constellation. For 16-QAM constellation, $\mathcal{S}_{iI}$ and $\mathcal{S}_{iQ}$ are of the form $\mathcal{S}_{iI} = \frac{1}{\sqrt{10}} \left\lbrace -3, -1, 1, 3 \right\rbrace \mbox{ and }\mathcal{S}_{iQ} = \frac{1}{\sqrt{10}} \{ -3\imath, -\imath, \imath, 3\imath \}.$ For this set-up, we obtain a two way partition of $\mathcal{S}_{iI}$ and $\mathcal{S}_{iQ}$ such that the minimum of the $d_{min}$ values of the sets in $\mathcal{A}$ is maximized. In particular, the optimal partition is obtained for different $P$ values. For each value of $P$, the optimal partition is obtained by using the corresponding $\alpha^{*}$ as given in the last column of Table \ref{table_for_scale_gmac}. The optimal partitions for all values of $P$ are found to be $\mathcal{S}^{1}_{iI} = \frac{1}{\sqrt{10}} \{ -3, 1 \}, \mathcal{S}^{2}_{iI} = \frac{1}{\sqrt{10}} \{ -\imath, 3\imath \},$ $\mathcal{S}^{1}_{iQ} = \frac{1}{\sqrt{10}} \{ -3, 1 \} \mbox{ and }\mathcal{S}^{2}_{iQ} = \frac{1}{\sqrt{10}} \{ -\imath, 3\imath \},$ which is the Ungerboeck partitioning. With this set partitioning, trellis code pairs based on TCM can be designed in order to transmit 2 bits for each user (for each user $m_{iX} = 1$ bit is transmitted along each dimension) using the 16-QAM constellation.

In the preceding sections, we have studied the advantage of the CPA scheme for two-user GMAC with no phase offsets in the channel. In the next section, we study the robustness of the CPA scheme for random phase offsets in the channel.
\section{CPA scheme for Two-User GMAC with Random Phase-Offsets}
\label{sec5}
In this section, we consider a two-user GMAC with random phase offsets introduced by the channel for both the users \cite{FAR}. Similar to the signal model in Section \ref{sec2}, the two users are equipped with constellations $\mathcal{S}_{1}$ and $\mathcal{S}_{2}$ of size $N_{1}$ and $N_{2}$, respectively. When User-$1$ and User-$2$ transmit symbols $\sqrt{P_{1}}x_{1}$ and $\sqrt{P_{2}}x_{2}$ simultaneously, the destination receives the symbol $y$ given by
\begin{equation}
\label{gmac_modeleq}
y = e^{\imath \theta_{1}}\sqrt{P_{1}} x_{1} + e^{\imath \theta_{2}}\sqrt{P_{2}} x_{2} + z, ~~ \mbox{where }  z \sim \mathcal{CSCG} \left(0, \sigma^{2} \right),
\end{equation}
and $\theta_{1}$, $\theta_{2}$ are i.i.d. random variables distributed uniformly over $(0, 2 \pi)$. We assume that only the destination has the knowledge of $\theta_{1}$ and $\theta_{2}$. If User-$i$ has the knowledge of $\theta_{i}$, then the phase offset can be compensated by each user which in turn results in the signal model discussed in Section \ref{sec2}. The CC capacity region for the above model can be computed along the similar lines of the model in Section \ref{sec2}. In addition to the steps needed to compute the mutual information values in Section \ref{sec2}, in this case, we have to take expectation of the mutual information values over $\theta_{1}$ and $\theta_{2}$. Hence, the CC sum capacity is given by $E_{\theta_{1}, \theta_{2}} \left[ I\left (\sqrt{P_{1}}x_{1} + \sqrt{P_{2}}x_{2} : y ~|~ \theta_{1}, \theta_{2} \right) \right]$, where $I \left(\sqrt{P_{1}}x_{1} + \sqrt{P_{2}}x_{2} : y ~|~ \theta_{1}, \theta_{2} \right)$ is given in \eqref{rp_mi}.

\begin{figure*}
\begin{equation}
I\left(\sqrt{P_{1}}x_{1} + \sqrt{P_{2}}x_{2} : y ~|~ \theta_{1}, \theta_{2}\right) = \mbox{log}_{2}(N_{1}N_{2}) - \nonumber
\end{equation}
\begin{equation}
\label{rp_mi}
\frac{1}{N_{1}N_{2}}\sum_{k_{1} = 0}^{N_{1} - 1}\sum_{k_{2} = 0}^{N_{2} - 1}E\left[\mbox{log}_{2}\left[ \frac{\sum_{i_{1} = 0}^{N_{1} - 1}\sum_{i_{2} = 0}^{N_{2} - 1} \mbox{exp}\left(- |e^{\imath \theta_{1}}\sqrt{P_{1}}(x_{1}(k_{1}) - x_{1}(i_{1})) + e^{\imath \theta_{2}}\sqrt{P_{2}}(x_{2}(k_{2}) - x_{2}(i_{2})) + z|^{2}/\sigma^{2}\right)}{\mbox{exp}\left(- |z|^{2}/\sigma^{2}\right)}\right] \right]
\end{equation}
\begin{eqnarray}
\label{thmeq_unqp}
Q_{p}(\bar{\alpha}) = E_{\theta_{1}, \theta_{2}} \left[ \sum_{k_{1} = 0}^{N_{1} - 1}\sum_{k_{2} = 0}^{N_{2} - 1}\mbox{log}_{2}\left[ \sum_{i_{1} = 0}^{N_{1} - 1}\sum_{i_{2} = 0}^{N_{2} - 1} \mbox{exp}\left(- |\mu_{1}(k_{1}, k_{2}, i_{1}, i_{2})|^{2}/2\sigma^{2}\right)\right]\right]
\end{eqnarray}
\begin{equation}
\label{mu_value}
\mu_{1}(k_{1}, k_{2}, i_{1}, i_{2}) = e^{\imath \theta_{1}} \left( \sqrt{P_{L}}x_{1}(k_{1}) - \sqrt{P_{L}}x_{1}(i_{1})\right) + e^{\imath \theta_{2}}\left(\sqrt{P_{S}}x_{2}(k_{2})
 - \sqrt{P_{S}}x_{2}(i_{2})\right)
\end{equation}
\hrule
\end{figure*}

\indent Note that the CC sum capacity is a function of the distance distribution (DD) of the sum constellation $\mathcal{S}_{sum}$ given by 
\begin{equation*}
\mathcal{S}_{sum} \triangleq \left\lbrace \sqrt{P_{1}} e^{\imath \theta_{1}} x_{1} + \sqrt{P_{2}} e^{\imath \theta_{2}} x_{2} ~ | ~ \forall ~ x_{1} \in \mathcal{S}_{1}, x_{2} \in \mathcal{S}_{2}\right\rbrace.
\end{equation*}
Despite having random phase offsets in the channel, the DD of $\mathcal{S}_{sum}$ can be changed by scaling the input symbols of one user relative to the other. With such a change in the DD of $\mathcal{S}_{sum}$, the CC sum capacity can be increased by choosing an appropriate $\alpha$. Towards that direction, we apply the CPA scheme (given in Section \ref{sec3}) for this channel. With the CPA scheme, the CC sum capacity is given by $E_{\theta_{1}, \theta_{2}} \left[ \sum_{\bar{\alpha} \in \Omega} I\left(\sqrt{(2-\bar{\alpha})P_{1}}x_{1} + \sqrt{\bar{\alpha} P_{2}}x_{2} : y ~|~ \theta_{1}, \theta_{2} \right)\right]$, where $\Omega = \{ \alpha, 2 - \alpha \}$. Since the CC sum capacity is a function of $\alpha$, we have to compute $\alpha$ such that $$E_{\theta_{1}, \theta_{2}} \left[ \sum_{\bar{\alpha} \in \Omega} I\left(\sqrt{P_{L}}x_{1} + \sqrt{P_{S}}x_{2} : y ~|~ \theta_{1}, \theta_{2} \right) \right]$$ is maximized, where $P_{L} = (2-\bar{\alpha})P_{1}$ and $P_{S} = \bar{\alpha} P_{2}$. In other words, we have to solve the following optimization problem,

\begin{small}
\begin{equation}
\label{rp_max_problem}
\alpha_{\small{\mbox{opt}}} = \argmax_{\alpha \in (0, 1]} E_{\theta_{1}, \theta_{2}} \left[ \sum_{\bar{\alpha} \in \Omega} I(\sqrt{P_{L}}x_{1} + \sqrt{P_{S}}x_{2} : y ~|~ \theta_{1}, \theta_{2}) \right].
\end{equation}
\end{small}

Note that the closed form expression of $E_{\theta_{1}, \theta_{2}} [I(\sqrt{P_{L}}x_{1} + \sqrt{P_{S}}x_{2} : y ~|~ \theta_{1}, \theta_{2})]$ is not available. Therefore, computing $\alpha_{\small{\mbox{opt}}}$ is not straightforward. On the similar lines of Theorem \ref{thm}, we use $\alpha^{*} = \mbox{arg} \min_{\alpha \in (0, 1]} \sum_{\bar{\alpha} \in \Omega}  Q_{p}(\bar{\alpha}),$ to obtain the appropriate values of $\alpha$, where $Q_{p}(\bar{\alpha})$ is given in \eqref{thmeq_unqp}, and $\mu_{1}(k_{1}, k_{2}, i_{1}, i_{2})$ is given in \eqref{mu_value}. Note that $Q_{p}(\bar{\alpha})$ is free from the random variable $z$. Hence, it is computationally easier to obtain $\alpha^{*}$.

\subsection{Numerical Results}

\begin{figure}[h]
\centering
\includegraphics[width=3in]{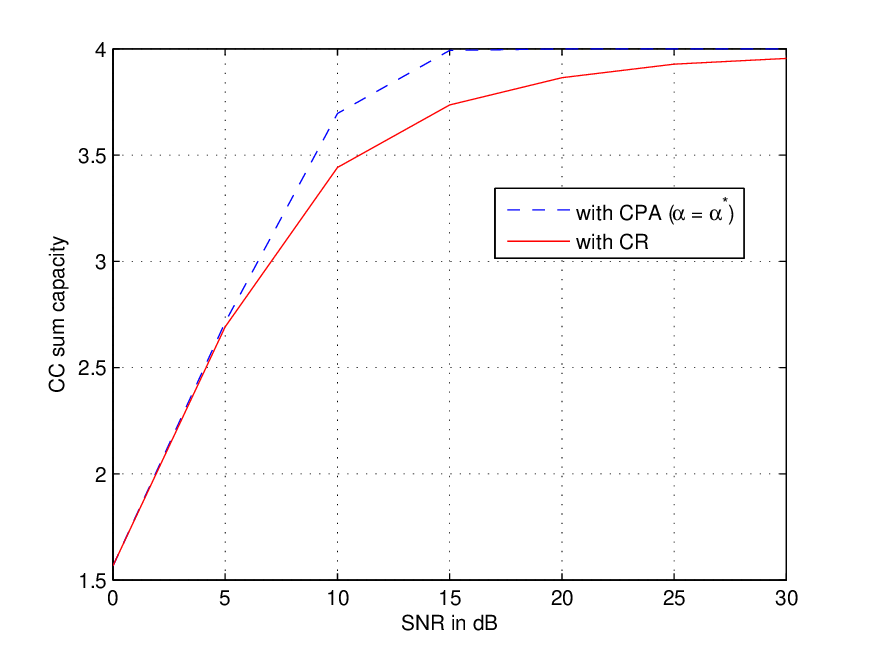}
\caption{CC sum capacity of QPSK with equal average power constraint and random phase offsets.}
\label{plot_gmac_rp_QPSK}
\end{figure}

\begin{figure}[h]
\centering
\includegraphics[width=3in]{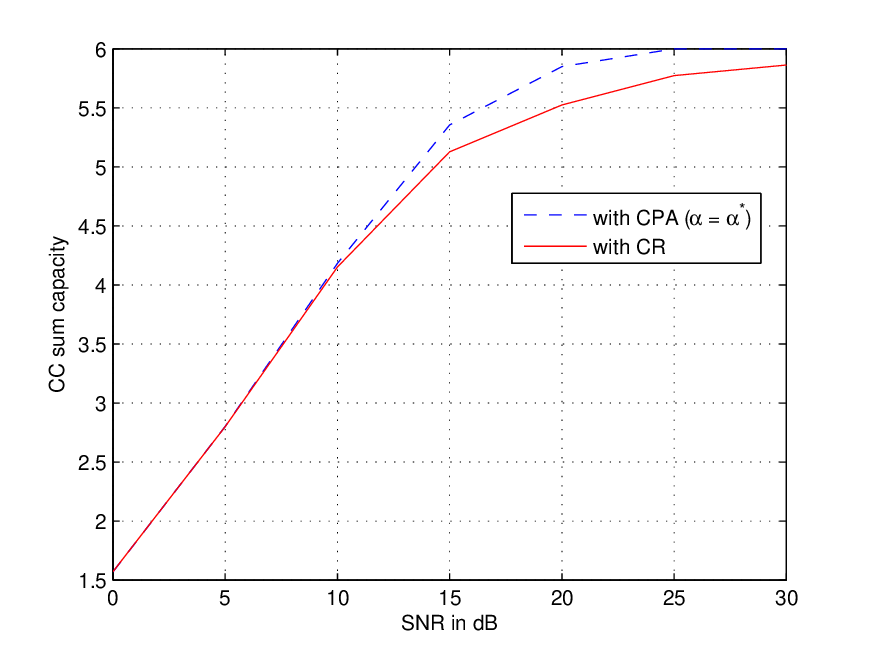}
\caption{CC sum capacity of 8-PSK with equal average power constraint and random phase offsets.}
\label{plot_gmac_rp_8_PSK}
\end{figure}

For the equal average power case (i.e., $\mbox{SNR} = P_{1} = P_{2}$ with $\sigma^2 = 1$), we compute the values of $\alpha^{*}$ for QPSK, 8-PSK and 8-QAM constellations. The corresponding values of $\alpha^{*}$ are listed in Table \ref{table_for_scale_gmac_rp}. The CC sum capacities for QPSK and 8-PSK are also provided in Fig. \ref{plot_gmac_rp_QPSK} and Fig. \ref{plot_gmac_rp_8_PSK}, respectively for the following two cases: ({\em i}) the CPA scheme with $\alpha = \alpha^{*}$, and ({\em ii}) the CR scheme. For the CR scheme, though the two users employ rotated constellations, the channel induced rotations (due to $\theta_{1}$ and $\theta_{2}$) make the effective angle of rotation a random variable. Therefore, the CR scheme for this channel corresponds to the CPA scheme with $\alpha = 1$. The figures show that the CPA scheme provides larger CC sum capacities than the CR scheme at high SNR values.

\begin{figure}[h]
\centering
\includegraphics[width=3in]{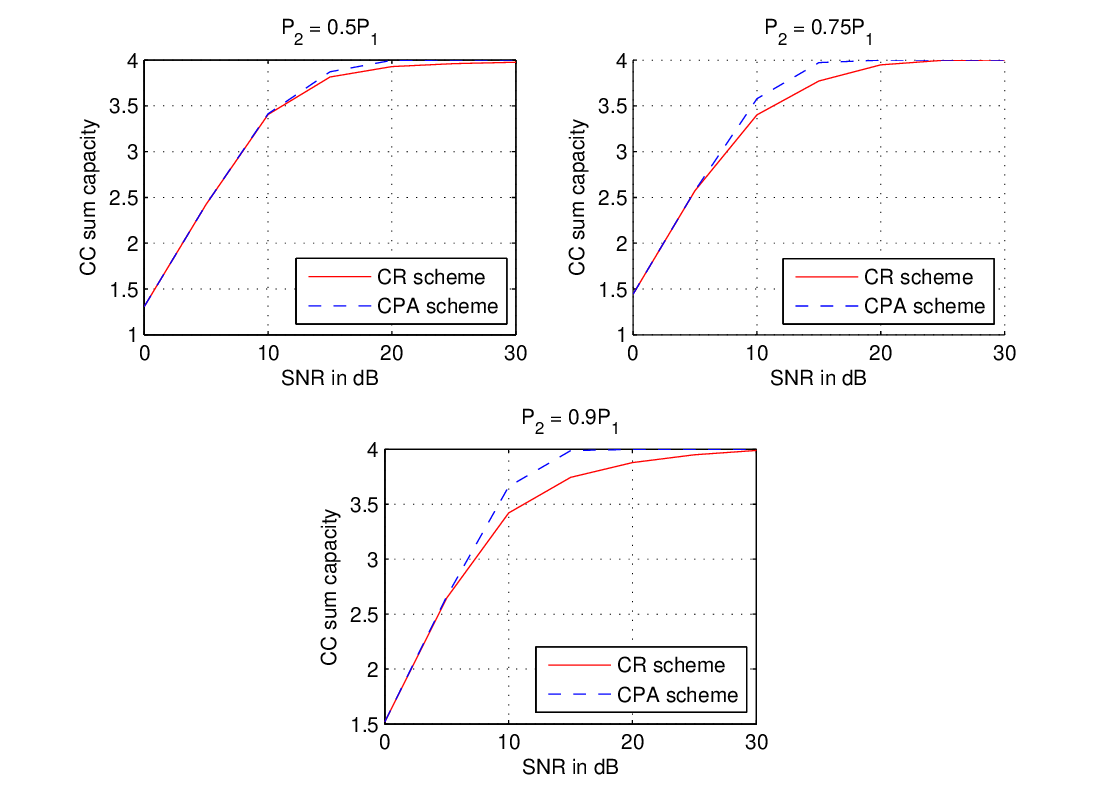}
\caption{CC sum capacities of QPSK with unequal average power constraint and random phase offsets.}
\label{uneqp_rp}
\end{figure}

\indent For the unequal average power case, we have computed $\alpha^{*}$  for the QPSK constellation and for several values of $\mbox{SNR} = \frac{P_{1}}{\sigma^{2}}$ (where $\sigma^{2} = 1$). We use the following three relations between $P_{1}$ and $P_{2}$: ({\em i}) $P_{2}$ = $0.5P_{1}$, ({\em ii}) $P_{2}$ = $0.75P_{1}$ and ({\em iii}) $P_{2}$ = $0.9P_{1}$. The corresponding CC sum capacities are presented in Fig. \ref{uneqp_rp} for the CPA and the CR scheme. From the figures, note that the CPA scheme provides increased CC sum capacities at moderate to high SNR values, especially when $P_{1}$ and $P_{2}$ are close. Further, the advantage of CPA is noticed to diminish when the power difference is more than $3$ dB.

\begin{center}
\begin{table*}
\caption{Numerically computed $\alpha^{*}$ with equal average power constraint and random phase offsets.}
\begin{center}
\begin{tabular}{|c|c|c|c|c|c|c|c|c|c|c|}
\hline SNR in dB & QPSK & 8-QAM & 8-PSK\\
\hline 0 & 0.86 & 0.97 & 0.75\\
\hline 5 & 0.54 & 0.97 & 0.72\\
\hline 10 & 0.38 & 0.66 & 0.62\\
\hline 15 & 0.32 & 0.45 & 0.54\\
\hline 20 & 0.30 & 0.13 & 0.16\\
\hline 25 & 0.12 & 0.12 & 0.15\\
\hline 30 & 0.04 & 0.10 & 0.13\\
\hline
\end{tabular}
\end{center}	 
\label{table_for_scale_gmac_rp}
\end{table*}
\end{center}

\section{Conclusion and Directions for Future Work}
\label{sec6}
The proposed CPA scheme can be useful for two-user fading MAC especially when the channel state information is available at the transmitters. For such a case, depending on the instantaneous channel states of the two users, we obtain one of the channel models discussed in this paper. Therefore, whenever the fading amplitudes of the two channels are approximately close, the CPA scheme will be beneficial. Hence, studying the impact of the CPA scheme for fading MAC is an interesting direction for future work. On the other hand, the CPA scheme can also be studied for GMAC with arbitrary number of users. However, large number of users in a NO-MA scheme is known to increase the decoding complexity at the destination \cite{MudV}. To avoid this increased complexity, the proposed results for two-user GMAC can be incorporated by the network schedulers, wherein instead of \emph{separating} all the users, the scheduler can \emph{separate} pairs of users, and then employ the CPA scheme between the two users in each pair.
\newpage

%

\begin{thebibliography}{1}
\bibitem{Ga}
R. Gallager, ``A Perspective on Multiaccess Channels," \emph{IEEE Trans. Information theory}, vol. 31, no. 02, pp. 124-142, Mar. 1985.
\bibitem{BiG}
E. Bigleiri, and L. Gyorfi, \emph{Multiple Access Channels: Theory and Practice}, IOS Press, Published in Cooperation with NATO Public Diplomacy Division, 2007.
\bibitem{As}
R. Ahlswede, ``Multi-Way Communication Channels," in Proc. of \emph{IEEE ISIT 1971}, Armenian, S.S.R, 1971.
\bibitem{Li}
H. Liao, ``A Coding Theorem for Multiple Access Communications," in Proc. of \emph{IEEE ISIT 1972}, Asilomar, CA, 1972.
\bibitem{RiU}
B. Rimoldi, and R. Urbanke, ``A Rate-Splitting Approach to the Gaussian Multiple-Access Channel,"  \emph{IEEE Trans. Information theory}, vol. 42, no. 02, pp. 364--375, Mar. 1996.

\bibitem{TsH1}
D. Tse, and S. Hanly, ``Multi-Access Fading Channels: Part I: Polymatroid Structure,
Optimal Resource Allocation and Throughput Capacities," \emph{IEEE Trans. Information theory}, vol.
44, no. 07, pp 2796--2815, Nov. 1998.

\bibitem{TsH2}
S.V. Hanly, and D. N. Tse, ``Multiaccess Fading Channels-Part II:Delay-Limited Capacities,"
\emph{IEEE Trans. Information theory}, vol. 44, no. 07, pp. 2816--2831, Nov. 1998.

\bibitem{CoT}
T. M. Cover, and J. A. Thomas, \emph{Elements of Information Theory}, 2nd
ed. Hoboken, NJ: Wiley, 2006.

\bibitem{MudV}
S. Verdu, \emph{Multiuser Detection}, Cambridge University Press, New York, 1998.

\bibitem{HYS}
H. Weingarten, Y. Steinberg, and S. Shamai, ``The Capacity Region of the Gaussian Multiple-Input Multiple-Output Broadcast Channel," \emph{IEEE Trans. Information theory}, vol. 52, no. 09, pp. 3936--3964, Sept. 2006.

\bibitem{ASNS}
A. Goldsmith, S. Jafar, N. Jindal, and S. Vishwanath, ``Capacity Limits of MIMO Channels," \emph{IEEE Journal on Selected Areas in Communication}, vol. 21, no. 05, pp. 684--702, Jun. 2003.

\bibitem{Ub}
G. Ungerboeck, ``Channel Coding with Multilevel/Phase Signals," \emph{IEEE Trans. Information theory}, vol. 28, no. 01, pp. 55--67, Jan. 1982.

\bibitem{FTL}
F. N. Brannstrom, T. M. Aulin, and L. K. Rasmussen, ``Constellation-Constrained Capacity for Trellis Code Multiple Access Systems," in Proc. of \emph{IEEE GLOBECOM 2001}, San Antonio, Texas, Nov. 2001.

\bibitem{FAR}
F. N. Brannstrom, Tor M. Aulin, and L. K. Rasmussen, ``Iterative Multi-User Detection of Trellis Code Multiple Access using a-posteriori Probabilities," in Proc. of \emph{IEEE ICC 2001}, Finland, Jun. 2001.

\bibitem{HaR_TIT}
J. Harshan, and B. S. Rajan, ``On Two-User Gaussian Multiple Access Channels with Finite Input Constellations," \emph{IEEE Trans. Information theory}, vol. 57, no. 03, pp. 1299--1327, Mar. 2011.

\bibitem{HaR_PIMRC}
J. Harshan, and B. S. Rajan, ``A Constellation Power Allocation Scheme for Two-User Gaussian MAC," in Proc. of \emph{IEEE PIMRC 2011}, Toronto, Canada, Sept. 2011.

\bibitem{SSA}
S. Chandrasekaran, Saif K. Mohammed, and A. Chockalingam, ``On the Capacity of Quantized Gaussian MAC Channels with Finite Input Alphabet," in Proc. of \emph{IEEE ICC 2011}, Kyoto, Japan, Jun 2011.

\bibitem{MWC}
M. Wang, W. Zeng, and C. Xiao, ``Linear Precoding for MIMO Multiple Access Channels with Finite Discrete Inputs," \emph{IEEE Trans. on Wireless Communications}, vol. 10, no. 11, pp. 3934--3941, Nov. 2011.

\bibitem{NaS}
N. Deshpande, and B. S. Rajan, ``Constellation Constrained Capacity of Two-User Broadcast Channels," in Proc. of \emph{IEEE GLOBECOM 2009}, Honolulu, Hawai, USA, Nov. 2009.

\bibitem{FrA}
F. Knabe, and A. Sezgin, ``Achievable Rates in Two-User Interference Channels with Finite Inputs and (very) Strong Interference," in Proc. of \emph{Asilomar Conference on Signals, Systems, and Computers 2010}. Also available in arXiv:1008.3035v1, Aug. 2010.

\bibitem{AbR}
G. Abhinav, and B. S. Rajan, ``Two-User Gaussian Interference Channel with Finite Constellation Input and FDMA," \emph{IEEE Trans. on Wireless Communications}, vol. 11, no. 07, pp. 2496--2507, Jul. 2012.

\bibitem{WMCJ}
W. Zeng, M. Wang, C. Xiao, and J. Lu, ``On the Power Allocation for Relay networks with Finite-Alphabet Constraints," in Proc. of \emph{IEEE GLOBECOM 2010}, Florida, USA, Dec. 2010.

\bibitem{ViR}
V. T Muralidharan, and B. S. Rajan, ``Bounds on the Achievable Rate for the Fading Relay Channel with Finite Input Constellations," available in arXiv:1102.4272v1, Feb. 2011.

\bibitem{WCMJ}
W. Zeng, C. Xiao, M. Wang, and J. Lu, ``Linear Precoding for Relay Networks with Finite-Alphabet Constraints," in Proc. of \emph{IEEE ICC 2011}, Jun. 2011. Also available in arXiv:1101.1345v2, Jan. 2011.

\bibitem{WCJ}
W. Zeng, C. Xiao, and J Lu, ``A Low-Complexity Design of Linear Precoding for MIMO Channels with Finite-Alphabet Inputs," \emph{IEEE Wireless Communications letters}, vol. 01, no. 01, pp. 38--41, Feb. 2012.

\bibitem{WXDZS}
Y. Wu,  C. Xiao, Z. Ding, Gao, X. Gao, and S. Jin, ``Linear Precoding for Finite Alphabet Signalling over MIMOME Wiretap Channels," \emph{IEEE Trans. Vehicular technology}, vol. 61, no. 06, pp. 2599--2612, Jul. 2012.

\bibitem{RaR}
G. D. Raghava, and B. S. Rajan, ``Secrecy Capacity of the Gaussian Wire-Tap Channel with Finite Complex Constellation Input," available in arXiv:1010.1163v1, Oct. 2010.

\bibitem{Eb}
E. Biglieri, \emph{Coding for Wireless Channels}, Springer-Verlag New York, Inc, 2005.
\end{thebibliography}
\end{document}